\documentclass[floatfix,aps,twocolumn,prb,superscriptaddress, ]{revtex4}
\usepackage{amsmath}
\usepackage{amssymb}
\usepackage{graphicx}
\usepackage{dcolumn}
\usepackage{natbib}
\usepackage{bm}
\usepackage{epsfig}

\begin{document}

\title{Disparities in the Josephson vortex state electrodynamics of high-$%
T_{c}~$cuprates}
\author{A.D. LaForge}
\email{alaforge@physics.ucsd.edu}
\affiliation{Department of Physics, University of California, San Diego, La Jolla,
California 92093, USA}
\author{W.J.~Padilla}
\altaffiliation{Present address: Department of Physics, Boston College, 140 Commonwealth Ave., Chestnut Hill, MA 02467, USA.}
\affiliation{Department of Physics, University of California, San Diego, La Jolla,
California 92093, USA}
\author{K.S.~Burch}
\altaffiliation{Present address: Los Alamos National Laboratory, MS K771, MPA-CINT, Los Alamos, New Mexico 87545, USA.}
\affiliation{Department of Physics, University of California, San Diego, La Jolla,
California 92093, USA}
\author{Z.Q. Li}
\affiliation{Department of Physics, University of California, San Diego, La Jolla,
California 92093, USA}
\author{S.V. Dordevic}
\affiliation{Department of Physics, The University of Akron, Akron, OH 44325, USA}
\author{Kouji Segawa}
\affiliation{Central Research Institute of the Electric Power Industry, Komae, Tokyo
201-8511, Japan}
\author{Yoichi Ando}
\affiliation{Central Research Institute of the Electric Power Industry, Komae, Tokyo
201-8511, Japan}
\author{D.N.~Basov}
\affiliation{Department of Physics, University of California, San Diego, La Jolla,
California 92093, USA}
\date{\today }

\begin{abstract}
We report on far infrared measurements of interplane conductivity for
underdoped single-crystal YBa$_{2}$Cu$_{3}$O$_{y}$ in magnetic field and
situate these new data within earlier work on two other high-T$_{c}$ cuprate
superconductors, La$_{2-x}$Sr$_{x}$CuO$_{4}$, and
Bi$_{2}$Sr$_{2}$CaCu$_{2}$O$_{8+d}$. The three systems have displayed apparently disparate
electrodynamic responses in the Josephson vortex state formed when magnetic field H is applied parallel to the CuO$_{2}$ planes. Specifically, there is discrepancy in the number and field dependence of longitudinal modes observed.   
 We compare and contrast these findings with
several models of the electrodynamics in the vortex state and suggest that most differences can be reconciled through considerations of the Josephson vortex lattice ground state as well as the c-axis and in-plane quasiparticle dissipations.
\end{abstract}

\maketitle

The superconducting vortex state, in which magnetic flux penetrates type II
superconductors in quantized vortices, has developed into a vast field of
theoretical and experimental research. Especially rich are the properties of
the vortex state in the high-temperature superconductors, where the CuO$_{2}$
planar structure and resulting anisotropic electronic structure introduce
fundamental topological differences between vortices produced by magnetic
fields oriented parallel and perpendicular to the CuO$_{2}$ planes.\cite
{Blatter-vortices-RMP1994} Probing the response of cuprate materials with
the $E$-vector of incoming radiation perpendicular to the CuO$_{2}$ planes
enables direct experimental access to many interesting and subtle features
of the vortex state.\cite
{Blatter-vortices-RMP1994,Golosovsky-1996,Kakeya-Kadowaki-BSCCO-vortex-modes-PRB2005,Basov-High-Tc-RMP2005}
The interlayer electrodynamics of cuprates in the superconducting state is
dominated by a resonance associated with coherent pair tunneling between the
planes, namely the Josephson plasma resonance (JPR). The JPR mode occurs in the
microwave range in Bi$_{2}$Sr$_{2}$CaCu$_{2}$O$_{8+d}$ (B2212) materials but lies in the far-IR region in YBa$_{2}$Cu$_{3}$O$_{y}$ (YBCO) and
La$_{2-x}$Sr$_{x}$CuO$_{4}$ (La214) due to the vastly different degree of anisotropy between these
compounds. Here we sum up new and previously published data on the JPR
response for the three families of cuprates. This analysis allows us to identify the key aspects of a comprehensive description of the Josephson vortex state in high-$T_{c}$ superconductors. Specifically, dissimilar features of the Josephson vortex electrodynamics can be reconciled by considering the role of both in-plane and $c$-axis dissipation following a recent theoretical treatment by Koshelev.\cite{Koshelev-preprint}

\begin{figure*}
\centering
   \includegraphics[width=6.5in, height=3.247in]{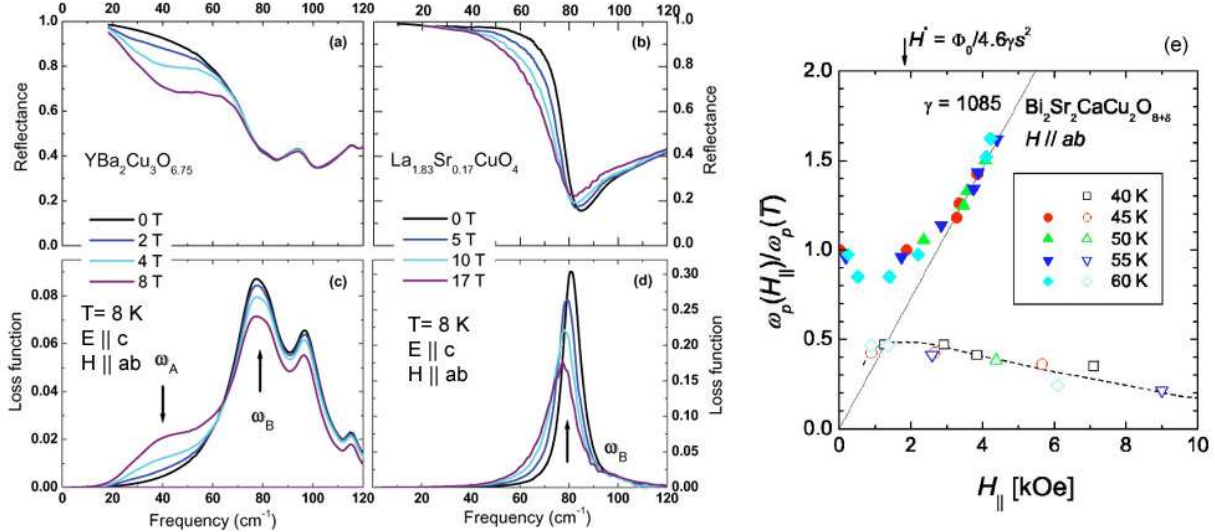}
     \caption{Comparison of electrodynamic response data for various families of cuprate superconductors. Raw reflectance spectra reveal a resonance feature below the Josephson plasma edge for YBCO (a) but not La214 (b). The loss function spectra show two longitudinal resonance modes for YBCO (c), but only one for La214 (d). A frequency-field phase diagram for Bi2212 (e) from ref. [\onlinecite{Kakeya-Kadowaki-BSCCO-vortex-modes-PRB2005}] displays two magnetoabsorption modes.}
     \label{Fig. 1}
\end{figure*}
Reflectance measurements were performed on high quality single crystals of
YBCO grown using the flux method\cite{Segawa-Ando-Transport} and assembled in a mosaic to form a reasonably large $ac$
face. Data were collected over wide ranges of
temperature (8-295 K), frequency (18-35 000 cm$^{-1}$),\cite{frequency-range} and magnetic field
(0-8 T, both parallel and perpendicular to the CuO$_{2}~$planes), all with
incident electric field polarized along the $c$ axis. Magnetic field
ratios were recorded as $R(\omega ,H)/R(\omega ,0~T)$, then multiplied by
zero field reflectance curves which had been normalized with an $in~situ~$gold
coating procedure to produce absolute reflectance as a function of field.%
\cite{Padilla-magnet-RSI2004,Homes-App-Optics1993} Data from far infrared to
UV were augmented with low and high frequency extrapolations and transformed
with the Kramers-Kronig relationship to obtain the complex conductivity $%
\hat{\sigma}(\omega )=\sigma _{1}\left( \omega \right) +i\sigma _{2}\left(
\omega \right)$ and dielectric function $%
\hat{\epsilon}(\omega )=\epsilon _{1}\left( \omega \right) +i\epsilon _{2}\left(
\omega \right).$

The raw reflectance data at $T=8~$K$~$are shown in Fig. 1(a) for several
values of magnetic field up to 8 T applied parallel to the CuO$_{2}$ planes in the Faraday configuration; Fig. 1(b) displays similar data for
La214 crystals up to 18 T from ref. [\onlinecite{SasaVortex}]. Below $T_{c}$
we observe the characteristic reflectance edge of the
JPR at frequency $\omega _{B}$.\cite
{Basov-c-axis-PRB1994,Shibauchi-pendepth-PRL1994,Dordevic-pendepth}
Application of magnetic field parallel to the CuO$_{2}$ planes impacts the
JPR in both systems, but in different ways. In YBa$_{2}$Cu$_{3}$O$_{6.75}$ a sharp dip in
reflectance appears at $\omega _{A}<\omega _{B}$ and moves to higher
energies with increasing magnetic field, but the JPR frequency is unchanged. Studies of more underdoped YBCO crystals have revealed an increase of the JPR frequency with field parallel to the CuO$_{2}$ planes. This was seen in the first optical studies of YBCO in field, at a doping of $y = 6.60$,\cite{Tajima-6.6-PRL2002} and has recently been verified in crystals with $y = 6.67$.\cite{to-be-published} In contrast, in the La214 data the JPR frequency $\omega _{B}$ softens with field, and the entire plasmon structure is weakened. For this system the qualitative trends are less sensitive to doping level.

The differences between the two systems are even more obvious upon
inspection of $\sigma _{1}\left( \omega \right) $, shown for YBCO in Fig. 2 and reported
elsewhere for La214.\cite{Dordevic-JPR-microscopy-PRL2003} The dip in
reflectance near $\omega_{A}$ is manifested in $\sigma _{1}\left( \omega \right) $ as a
transverse resonance which hardens and gains spectral weight in field. For
La214 no such resonance is observed, and $\sigma _{1}\left( \omega \right) $
exhibits no field-induced peak below the phonon range. Fields applied parallel to the $c$ axis (not shown) do not introduce a resonance in either system.

In order to facillitate direct comparison between the above IR results and
microwave data for Bi2212\cite{Kakeya-Kadowaki-BSCCO-vortex-modes-PRB2005}
it is instructive to turn to the spectra of the loss function, defined as -$Im(1/\hat{\varepsilon}(\omega ))$ and shown in Figs. 1(c) and 1(d).
The loss function spectra uncover the longitudinal modes in a system's
response, and thus can be related to the microwave magnetoabsorption
features in Fig. 1(e). The work in ref. [\onlinecite{Kakeya-Kadowaki-BSCCO-vortex-modes-PRB2005}] focused on an underdoped crystal with transition temperature $T_{c}$ = 70 K, but all trends were observed at optimal doping as well. The frequency-field diagram for Bi2212 displays two
resonances: one appears only at higher temperatures and hardens linearly with
field as a dense vortex lattice is formed;\cite{Ichioka-vortex-lattice-PRB1995} the other resonance, visible
at low temperature and nonzero fields, softens with magnetic field. This result differs distinctly
from that of the other systems; La214 supports only one sharp longitudinal mode, and its peak frequency $\omega _{B}$ decreases with field. In YBCO the JPR peak frequency 
$\omega _{B}$ is field independent or weakly increasing, and the linewidth is broader. Furthermore, both modes in YBCO are
sharpest at low temperature, with no evidence of the additional temperature scale seen
in Bi2212. The closest agreement between the data sets lies in the
lower-frequency modes of YBCO (labeled as $\omega _{A}$) and Bi2212. Both
are too weak to be resolved at the lowest fields and have little frequency
dependence in modest fields. At the outset, the electromagnetic responses of
the three systems appear to be quite distinct and without a common pattern; thus, the task of finding a universal
explanation has not been straightforward.

Many theoretical models have been proposed to explain the low-frequency
infrared and microwave properties of the layered high-$T_{c}$ superconductors.
 Older theories\cite
{Bulaevskii-parallel-field-PRL1991,Bulaevskii-c-axis-field-PRL1996,Koshelev-disorder-vortices-PRB1996}
have accurately described elements of the experimental data for individual
families of cuprates but have not sufficiently accounted for the differences in resonance behaviors
from family to family displayed in Fig. 1. Discussion below outlines a series of developments which form a coherent explanation of these disparities. A classical description of Josephson vortex oscillation presented by Tachiki, Koyama, and Takahashi (TKT)\cite{TKT} marks a good starting point for approaching this problem. This model
focuses on the Lorentz coupling between a $c$-axis polarized AC electric
field and a Josephson vortex lattice oriented parallel to the CuO$_{2}$
planes. In the presence of vortex pinning and viscosity, the electric
field will drive a vortex resonance that is visible as a dip in the raw
reflectance data, provided the vortex mass is set to a finite value. Approximations for low magnetic fields and frequencies allow the authors to neglect details of the vortex lattice configuration and reach a tractable analytic solution. 

The TKT theory affords insight into the influence of the vortex dynamical parameters and yields a good fit to the experimental data. However, application of this theory to the systems considered here relies on assumptions which may be invalid. First, the TKT model will produce a new field-induced resonance only if large effective
mass is assigned to Josephson vortices. Then within this framework one has
to assume massive vortices in YBCO and much lighter ones in La214, an
unlikely premise given the similarities in the zero-field response between
the two systems. Second, it is likely that the approximations for low frequencies
and fields place the features under consideration outside
of the physcially meaningful parameter space. 

\begin{figure}
\centering
   \includegraphics[width=3.375in, height=2.597in]{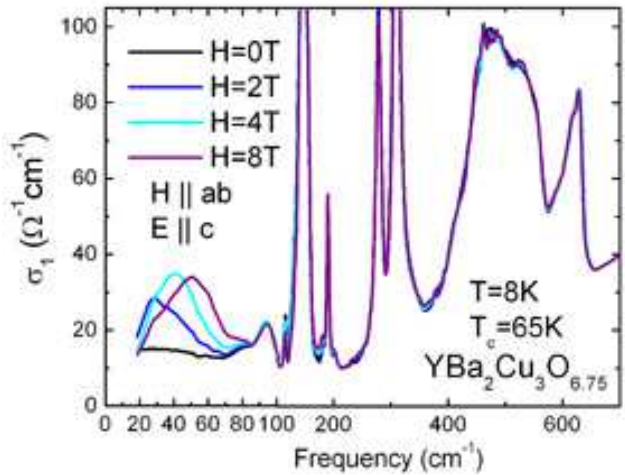}
     \caption{Optical conductivity of YBa$_{2}$Cu$_{3}$O$_{6.75}$ at 8 K for magnetic fields oriented 
parallel to the CuO$_{2}$ planes.}
     \label{Fig. 2}
\end{figure}

Another model by van der Marel and Tsvetkov (vdM/T)\cite{vdM-bilayer-Czech,Dulic-vdM-magnet-PRL2002,Pimenov-mag-SmLSCO-2001}
considers the effect of magnetic field upon interlayer Josephson coupling.
In this picture, a fraction of interlayer junctions are penetrated by
vortices in a superlattice structure, resulting in a renormalized JPR
frequency for those layers. The out-of-phase oscillation of charge in the
differing junctions then produces a transverse resonance which is observed
as a peak in $\sigma _{1}\left( \omega \right) $. This model yields a good fit to the present YBCO data with few free parameters,\cite
{to-be-published} and has an excellent track record in describing far-infrared resonances in a variety of systems with multilayer geometries.\cite{Dulic-vdM-magnet-PRL2002,Pimenov-mag-SmLSCO-2001}  The strength and versatility of the approach stem from its phenomenological handling of the modification of interlayer Josephson couplings. When augmented with detailed calculations of the Josephson vortex superstructure (discussed below) the vdM/T framework provides a qualitative account of differences between magneto-optics data in YBCO and La214 compounds. 

It is imperative to turn to the results of vortex lattice calculations in order to analyze on the same footing the in-field JPR response of different families of cuprates. Recent studies\cite{Koshelev-ground-states-2006,Nonomura-energy-landscape-PRB2006} of the Josephson vortex lattice ground state define the critical field scale as $H_{cr}=\Phi_{0}/2 \pi \gamma s^{2}$, where $\Phi_{0}$ is the magnetic flux quantum, $\gamma$ is the anisotropy parameter, and $s$ is the interlayer distance. For high fields the Josephson vortices fill every layer to form a dense lattice, but upon lowering to $H = H_{cr}$ it becomes favorable for each pair of layers containing vortices to be separated by an empty layer. As the field is further decreased there is a complicated series of first order transitions between configurations with varying spacings, until a dilute lattice is eventually formed for $H << H_{cr}$. For Bi2212, $H_{cr} \approx 0.39$ T, while for YBCO and La214 the field scales are 23 T and 93 T,
respectively. The limitations on experimentally available magnetic
field strength then place each sample in Fig. 1 in a different field regime. In the
case of La214, the maximum field experimentally available (17 T) is less than a quarter of $H_{cr}$, so
the vortex structure is still dilute. Thus, the data for La214 would be more
accurately compared to the far left side of Fig. 1(e), where the upper
frequency mode softens with field and the lower frequency mode is not yet
observed. For YBCO, however, the low-frequency loss function peak is first resolved just below $%
H_{cr}/2$, as was observed for Bi2212. The broad onset of this feature in YBCO at fields as low as $H_{cr}/10$ has not yet been reconciled with the single-peaked spectra of La214. Only for Bi2212, which boasts an anisotropy 50-100 times as large as that of YBCO or La214, is the dense
vortex lattice limit reached. 

\begin{figure}
\centering
   \includegraphics[width=3.375in, height=2.389in]{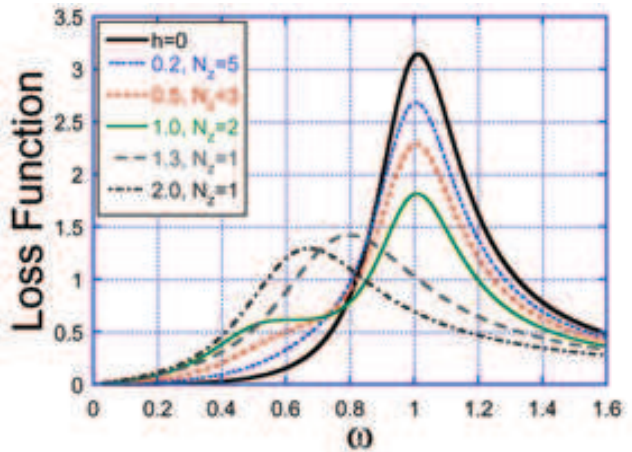}
     \caption{Theoretical loss function predicted by Koshelev model\cite{Koshelev-preprint} for a system with high dissipation $(\nu_{c}=0.32, \nu_{ab}=6.0$) in a static magnetic field $h=H/H_{cr}$ (see text). Values of $N_{z}$ refer to the number of layers between vortices.}
     \label{Fig. 3}
\end{figure}

Understanding of the Josephson vortex state electrodynamic response has been further advanced by the inclusion of another set of key parameters, the $c$-axis and in-plane dissipation values. Recently the equations describing phase dynamics in a layered superconductor
in parallel field have been solved numerically by Koshelev,\cite{Koshelev-preprint} yielding a
solution for the complex dielectric function $\hat{\varepsilon}\left( \omega \right)$ which is valid for
all frequencies and fields. This work begins with the coupled equations for the phase difference and magnetic field in the absence of charging effects\cite{Artemenko-phase-eq-JETPLett1997,Koshelev-phase-eq-PRB2001} and solves the static and dynamic phase equations in turn. This description takes into account the vortex lattice configuration discussed above, and
depends strongly upon both the in-plane and $c$-axis dissipation parameters, $\nu_{ab}=4 \pi \sigma_{c}/\epsilon_{c} \omega_{p}$ and $\nu_{c}=4 \pi \sigma_{ab} \lambda_{ab}^{2} \omega_{p}/c^{2}$, which scale roughly as the inverse of the anisotropy.
Also critical is the frequency dependence of their relative strengths. Such an approach provides a natural pathway for addressing the differences among cuprate families, and indeed many observed features are reproduced by the theory. For low
values of the dissipation parameters (typical of those measured in Bi2212), the model matches the field dependence of the two modes measured in that system. And for high dissipation, as realized in underdoped
YBCO, fields below $H_{cr }$ generate the observed depletion of the main loss function peak and
introduce a low-frequency mode, shown in Fig. 3. The model also exhibits a finite resonance in $\sigma _{1}\left( \omega \right) $ which hardens with magnetic field, in agreement with experimental observations.  

The reliance of this method upon the quasiparticle dissipation initiates a comparison across cuprate families. It is known, for example, that the DC conductivities along the $c$ axis of Bi2212 and YBCO can differ by three orders of magnitude.\cite%
{Thomas-Gough-Condmat2000} Also, the infrared/microwave data for YBCO reveal both a wider JPR linewidth and a stronger frequency dependence of the in-plane optical conductivity than is observed for La214.\cite{Dumm-LSCO-PRL2002,Hwang-orthoII-PRB2006,Bonn-quasiparticle-PRB1999}  This model, then, could be exposing the sensitivity of the JPR to 
these properties. For completeness, we briefly mention two other structural differences which could contribute to disparities: pinning and layeredness. The CuO chain structure and twin
boundaries, which are present only in YBCO, have been shown to affect
properties of vortex pinning\cite
{Gyorgy-twin-boundary-pinning-APL1990,Herbsommer-twin-boundary-pinning-PRB2000} and may in turn influence the vortex resonance spectra. And of the three systems discussed here, only La214 is single-layered, while YBCO and Bi2212 have 2 and 3 layers, respectively. This factor could affect the vortex lattice ground state configuration. 

We have shown that apparent disparities exist in the Josephson vortex state electrodynamic response of several families of cuprate superconductors. After examining proposed theoretical models we can conclude that the differences originate not in variations of vortex mass, but in anisotropy and dissipation. The description proposed by Koshelev\cite{Koshelev-preprint} represents a significant step towards a coherent understanding of the interlayer response of the Josephson vortex state. Future spectroscopic measurements which expand the experimental phase diagram with higher magnetic fields and lower frequencies\cite{Bonn-microwave-PRL2003} should further elucidate this subject.
\bigskip

We thank A. E. Koshelev for illuminating discussions and for sharing his drafts prior to publication. This research was supported by the United States Department of Energy and
the National Science Foundation. The work done at CRIEPI was supported by
the Grant-in-Aid for Science provided by the Japan Society for the Promotion
of Science.


\begin{thebibliography}{99}
\bibitem{frequency-range} Reflectance measurements were performed in magnetic field for frequencies up to 700 cm$^{-1}$, above which no field-induced changes were detected.
\end{thebibliography}
\end{document}